\documentclass[twocolumn]{aastex631}

\shorttitle{{\sc HIFlow}}
\shortauthors{Hassan et al.}
\graphicspath{{./}{figures/}}

\begin{document}

\title{{\sc HIFlow}: Generating Diverse HI Maps and Inferring Cosmology while Marginalizing over Astrophysics using Normalizing Flows}

\author[0000-0002-1050-7572]{Sultan Hassan}
\affiliation{Center for Computational Astrophysics, Flatiron Institute, 162 5th Ave, New York, NY, 10010, USA}\affiliation{Department of Physics \& Astronomy, University of the Western Cape, Cape Town 7535,
South Africa}

\author[0000-0002-4816-0455]{Francisco Villaescusa-Navarro}\affiliation{Department of Astrophysical Sciences, Princeton University, Peyton Hall, Princeton, NJ, 08544, USA}
\affiliation{Center for Computational Astrophysics, Flatiron Institute, 162 5th Ave, New York, NY, 10010, USA}

\author{Benjamin Wandelt}
\affiliation{Sorbonne Universite, CNRS, UMR 7095, Institut d’Astrophysique de Paris, 98 bis boulevard Arago, 75014 Paris, France}
\affiliation{Center for Computational Astrophysics, Flatiron Institute, 162 5th Ave, New York, NY, 10010, USA}

\author{David N. Spergel}
\affiliation{Center for Computational Astrophysics, Flatiron Institute, 162 5th Ave, New York, NY, 10010, USA}\affiliation{Department of Astrophysical Sciences, Princeton University, Peyton Hall, Princeton, NJ, 08544, USA}

\author[0000-0001-5769-4945]{Daniel Angl\'es-Alc\'azar}\affiliation{Department of Physics, University of Connecticut, 196 Auditorium Road, U-3046, Storrs, CT, 06269, USA}\affiliation{Center for Computational Astrophysics, Flatiron Institute, 162 5th Ave, New York, NY, 10010, USA}

\author{Shy Genel}
\affiliation{Center for Computational Astrophysics, Flatiron Institute, 162 5th Ave, New York, NY, 10010, USA}
\affiliation{Columbia Astrophysics Laboratory, Columbia University, New York, NY, 10027, USA}

\author{Miles Cranmer}\affiliation{Department of Astrophysical Sciences, Princeton University, Peyton Hall, Princeton, NJ, 08544, USA}

\author{Greg L. Bryan}
\affiliation{ Department of Astronomy, Columbia University, 550 West 120th Street, New York, NY 10027, USA}\affiliation{Center for Computational Astrophysics, Flatiron Institute, 162 5th Ave, New York, NY, 10010, USA}

\author{Romeel Dav\'e}
\affiliation{
Institute for Astronomy, University of Edinburgh, Royal Observatory, Edinburgh EH9 3HJ, UK}\affiliation{Department of Physics \& Astronomy, University of the Western Cape, Cape Town 7535,
South Africa}\affiliation{
South African Astronomical Observatories, Observatory, Cape Town 7925, South Africa}

\author{Rachel S. Somerville}\affiliation{Center for Computational Astrophysics, Flatiron Institute, 162 5th Ave, New York, NY, 10010, USA}

\author{Michael Eickenberg}\affiliation{Center for Computational Mathematics, Flatiron Institute, 162 5th Ave, New York, NY, 10010, USA}

\author{Desika Narayanan}
\affiliation{Department of Astronomy, University of Florida, Gainesville, FL, USA}
\affiliation{University of Florida Informatics Institute, 432 Newell Drive, CISE Bldg E251, Gainesville, FL, USA}

\author{Shirley Ho}
\affiliation{Center for Computational Astrophysics, Flatiron Institute, 162 5th Ave, New York, NY, 10010, USA}\affiliation{Department of Astrophysical Sciences, Princeton University, Peyton Hall, Princeton, NJ, 08544, USA}\affiliation{Department of Physics, Carnegie Mellon University, Pittsburgh, PA 15213, USA}

\author[0000-0001-5957-0719]{Sambatra Andrianomena}
\affiliation{
South African Radio Astronomy Observatory (SARAO), Black River Park, Observatory, Cape Town, 7925, South Africa}\affiliation{Department of Physics \& Astronomy, University of the Western Cape, Cape Town 7535,
South Africa}



\begin{abstract}
 A wealth of cosmological and astrophysical information is expected from many ongoing and upcoming large-scale surveys. It is crucial to prepare for these surveys now and develop tools that can efficiently extract most information. We present {\sc HIFlow}: a fast generative model of the neutral hydrogen (HI) maps that is conditioned only on cosmology ($\Omega_{m}$ and $\sigma_{8}$) and designed using a class of normalizing flow models, the Masked Autoregressive Flow (MAF). {\sc HIFlow} is trained on the state-of-the-art simulations from the Cosmology and Astrophysics with MachinE Learning Simulations (CAMELS) project. {\sc HIFlow} has the ability to generate realistic diverse maps  without explicitly incorporating the expected 2D maps structure into the flow as an inductive bias. We find that {\sc HIFlow} is able to reproduce the CAMELS average and standard deviation HI power spectrum (Pk) within a factor of $\lesssim$ 2, scoring a very high $R^{2} > 90\%$. By inverting the flow, {\sc HIFlow} provides a tractable high-dimensional likelihood for efficient parameter inference. We show that the conditional {\sc HIFlow} on cosmology is successfully able to marginalize over astrophysics at the field level, regardless of the stellar and AGN feedback strengths. This new tool represents a first step toward a more powerful parameter inference, maximizing the scientific return of future HI surveys, and opening a new avenue to minimize the loss of complex information due to data compression down to summary statistics.

\end{abstract}


\section{Introduction} \label{sec:intro}

Extracting the maximum amount of cosmological and astrophysical information remains a challenge in upcoming large-scale surveys such as the Square Kilometer Array ~\citep[SKA,][]{2013ExA....36..235M}, the Hydrogen Epoch of Reionization Array~\citep[HERA,][]{2017PASP..129d5001D},  the Low Frequency Array~\citep[LOFAR,][]{2013A&A...556A...2V}, the Vera C. Rubin Observatory Legacy Survey of Space and Time~\citep[LSST,][]{2019ApJ...873..111I}, Nancy Grace Roman Space Telescope~\citep[Roman,][]{2015arXiv150303757S}, Spectro-Photometer for the History of the Universe, Epoch of Reionization, and Ices Explorer~\citep[SPHEREx,][]{2014arXiv1412.4872D}, and  Euclid~\citep[][]{2016SPIE.9904E..0OR}.  Particularly, evaluating the exact likelihood of the expected high-dimensional data sets from these surveys remains intractable. Because of the large memory requirements associated with the upcoming data sets, many analyses use summary statistics, which in many cases (such as the commonly used power spectrum) results in throwing away a large amount of information.  Some recent works have presented methods to search for the optimal summary statistic which can successfully capture the non-gaussian complex features, such as the information maximising neural networks~\citep[IMNNs,][]{2018PhRvD..97h3004C} and wavelet scattering transform~\citep{2011arXiv1101.2286M}, to reduce dimensions while minimizing loss of information. While these methods show different levels of success, a natural way  to prevent loss of information is to perform inference at the field level. 

Convolutional Neural networks (CNNs) have been very successful in extracting information from high-dimensional data sets by capturing non-Gaussian features. A few examples of the successful use of CNNs include: constraining cosmology and astrophysics~\citep{2020MNRAS.494.5761H, baryons_marginalization, Robust_marginalization}, identifying sources driving cosmic reionization~\citep{2019MNRAS.483.2524H}, constraining the reionization history~\citep{2020MNRAS.494..600M}, learning galaxy properties from 21 cm lightcones~\citep{2021arXiv210700018P}, recovering astrophysical parameters~\citep{2019MNRAS.484..282G}, painting HI on the matter field from N-body simulations~\citep{2021ApJ...916...42W}, removing astrophysical effects \citep{Pablo_2020} and providing optimal summary statistics for simulation-based inference~\citep{2021arXiv210503344Z}. 

However, preparing these large-scale data sets requires running thousands of cosmological volumes using state-of-the-art hydrodynamic galaxy formation simulations by varying the cosmological and astrophysical parameters, which comes with a large computational expense. In addition, exploring the full parameter space controlling the astrophysical and cosmological observables is challenging. For instance, the state-of-the-art Cosmology and Astrophysics with MachinE Learning Simulations~\citep[CAMELS,][]{2021ApJ...915...71V} project, which is the largest data set designed to train machine learning models, provides only 1,000 simulations per subgrid model, that includes all the different recipes of modelling stellar and AGN feedback below the resolution limit, for exploring its 6-dimensional parameter space. 
Furthermore, linking these simulations directly to statistical tools, such as {\sc emcee}~\citep{2013PASP..125..306F} or {\sc pydelfi}~\citep{2019MNRAS.488.4440A}, to perform inference is beyond the reach of current computing capability. A powerful alternative approach is directly learn a tractable likelihood from datasets or simulations to perform parameter inference with a minimal cost. This is the goal of this paper.

Currently, there are several competing machine-learning techniques to generate new examples of large scale data sets. This includes Generative Adversarial Networks~\citep[][GANs,]{2014arXiv1406.2661G}, Variational AutoEncoder~\citep[VAE,][]{2013arXiv1312.6114K}, and Normalizing Flows~\citep[NF,][]{2015arXiv150505770J,2014arXiv1410.8516D}. Generating new {\it diverse} examples is crucial to ensure capturing the wide range of features present in the training sample. One common problem with GANs is called mode collapse, in which the generator always produces the same realization of the output. However,  the advantages of using NF over other methods is the ability to learn the exact likelihood function to perform either inference or generate new diverse examples by inverting the flow transformations. NF methods have been very successful in generating random cosmological fields~\citep{2021arXiv210512024R}, simulating galaxy images~\citep{2021MNRAS.504.5543L}, performing likelihood-free inference~\citep[e.g.][]{2019MNRAS.488.4440A}, and modelling  Color-Magnitude diagrams~\citep{2019arXiv190808045C}. NF attempts to learn the mapping between a standard Gaussian field and the more complex density distribution of the observable (in this case the HI maps). Once the mapping is found, new examples can be sampled simply from the initial Gaussian field. This is, in fact, somewhat conceptually similar to the flow within the most sophisticated galaxy formation and cosmological simulations. Most simulations in astrophysics and cosmology apply a series of recipes (i.e. to evolve the density and form stars) in order to transform an initial Gaussian distribution (i.e. the density field) to a complex nonlinear observable (e.g., large scale structure, reionization morphology, cosmic evolution of star formation). Similar to galaxy formation and cosmological simulations, NF models is able to generate new diverse examples for the same set of parameters, and hence they naturally capture cosmic variance effects.

In this paper, we present {\sc HIFlow}: a fast generative model of the neutral hydrogen (HI) maps by the end of reionization at z$\sim$6. We choose the HI fields since many of the upcoming future surveys aim to map out the HI distribution in the early Universe to trace the large-scale structure. To train our emulator, we use the HI maps that are generated in a similar way as described in the {\sc CAMELS} Multifield Dataset\footnote{\url{https://camels-multifield-dataset.readthedocs.io}} \citep[CMD,][]{CMD} but at z$=$6 with a lower resolution. The state-of-the-art CAMELS simulation contains thousands of HI maps generated using SIMBA~\citep{2019MNRAS.486.2827D} and IllustrisTNG~\citep{2017MNRAS.465.3291W,2018MNRAS.473.4077P} simulations. We here focus on the HI maps generated using the {\sc IllustrisTNG} simulations. We first train unconditional {\sc HIFlow} and test its performance by comparing with maps from CAMELS in terms of several summary statistics such as the probability density functions and the power spectrum. We next train a conditional {\sc HIFlow} on the two cosmological parameters, namely $\Omega_{m}$ and $\sigma_{8}$, and validate the results at the power spectrum level.  We show several examples for posterior distributions using the conditional {\sc HIFlow}. We finally demonstrate the ability of {\sc HIFlow} to perform cosmological inference while marginalizing over astrophysics at the field level.

This paper is organized as follows: We briefly discuss the simulations in \S\ref{sec:sims} and present the NF method used in~\S\ref{sec:maf}. The unconditional and conditional {\sc HIFlow} results are presented in \S\ref{sec:uhiflow} and \S\ref{sec:chiflow}, respectively.  We show several examples of how to perform efficient inference with {\sc HIFlow} in~\S\ref{sec:inf}, and marginalize over astrophysics in~\S\ref{sec:inf_com}. We summarize and make our concluding remarks in \S\ref{sec:con}.

\section{Simulations}\label{sec:sims}
We use simulations from the CAMELS project, that have been recently introduced in~\citet{2021ApJ...915...71V}. Here, we briefly describe CAMELS and refer the reader to~\citet{2021ApJ...915...71V} for further details on the different simulations and datasets. CAMELS is a suite of thousands of simulations run with state-of-the-art cosmological hydrodynamic galaxy formation models, namely {\sc Simba}~\citep{2019MNRAS.486.2827D} and {\sc IllustrisTNG}~\citep{2017MNRAS.465.3291W,2018MNRAS.473.4077P}, by varying two cosmological parameters ($\Omega_{m}$ and $\sigma_{8}$) and 
four other parameters that modify the strength of stellar feedback ($A_{\rm SN1}$, $A_{\rm SN2}$) and black hole feedback ($A_{\rm AGN1}$, $A_{\rm AGN2}$) relative to the original {\sc IllustrisTNG} and {\sc Simba} simulations.  
 
 We focus our analysis on two CAMELS sets, namely, the 1 parameter (1P) set which varies a single parameter at a time with the same initial seed number, and the Latin-hyper-cube (LH) set, which is a set of 1,000 simulations that explores this six-dimensional parameter space with uniform prior ranges defined as: $\Omega_{m} \in (0.1,0.5),\, \sigma_{8} \in (0.6,1.0),\, A_{\rm SN1} \in (0.25,4.0),\, A_{\rm AGN1} \in (0.25,4.0),\, A_{\rm SN2} \in (0.5,2.0)$, and $A_{\rm AGN2} \in (0.5,2.0)$, with different initial seeds. 

\section{Masked Autoregressive Flow}\label{sec:maf}
{\sc HIFlow} is designed following closely the method presented in~\citet{2017arXiv170507057P,2015arXiv150203509G}.  Normalizing flows (NF) are a class of generative models, which allow for tractable and efficient density estimation. The core principle of NF is the change-of-variable formula, which constructs a mapping ($f$) between a base distribution ($\pi_{u}({\bf u})$, usually a Gaussian) and a more complex distribution $p({\bf x})$ (i.e. the observable). Having obtained $f$, a new example of $\bf{x}$ can be generated using $\bf{x}=f(\bf{u)}$, where $\bf{u}$ is randomly drawn from the base distribution (${\bf{u}}\sim\pi_{u}({\bf u})$). This transformation ($f$) is required to be invertible as well as differentiable so that the target density $p(\bf{x})$ can be exactly evaluated as 

\begin{equation}\label{eq:eq1}
    p(\textbf{x}) = \pi_{u} (f^{-1}(\textbf{x})) \left|\det\left(\frac{\partial f^{-1}}{\partial \textbf{x}}\right)\right|\, ,
\end{equation}
where a tractable Jacobian is needed for easy computation of the determinant.
For instance, if $f$ is a series of transformations (e.g. f$_{i}$=$(\exp(\alpha_{i}))^{2}$), then it is straightforward to find the determinant of its Jacobian as follows:
\begin{equation}
    \left|\det\left(\frac{\partial f^{-1}}{\partial \textbf{x}}\right)\right| = \prod^{N}_{i=1}   \left|\det\left(\frac{\partial f_{i}^{-1}}{\partial \textbf{x}}\right)\right| = \exp\left( -2 \sum_{i}\alpha_{i} \right).
\end{equation}

\begin{figure}
\centering
\includegraphics[scale=0.5]{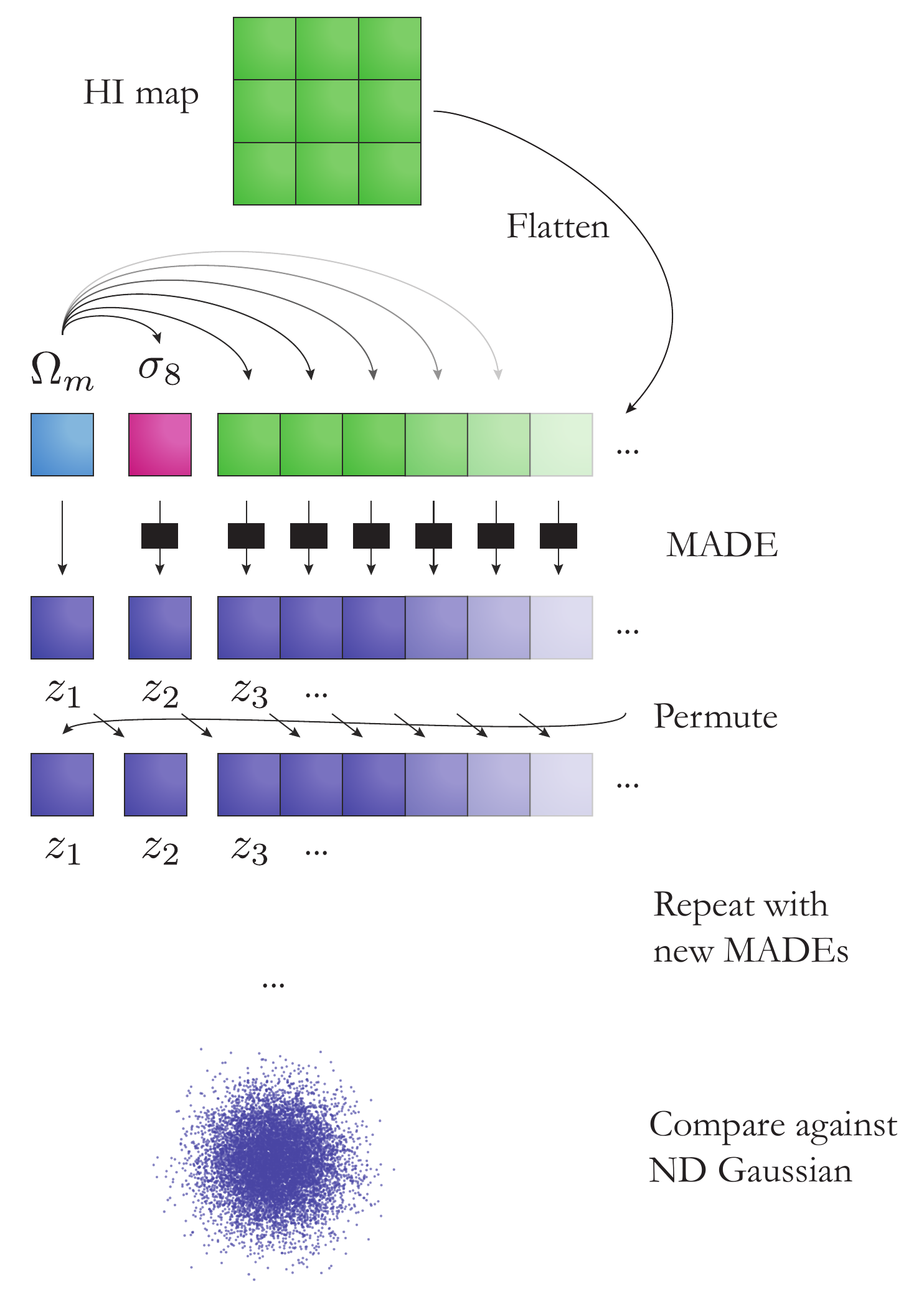}
    \caption{Diagram of the inference scheme by the {\it conditional} {\sc HIFlow} on the cosmological parameters ($\Omega_{m}$ and $\sigma_{8}$). Arrows indicate conditional dependence between variables, which constructs the flow between all MADEs in order to produce the joint density from the N-dimensional Gaussian distribution. The $z$ variables denote the latent spaces in MADEs.}
    \label{fig:graph}
\end{figure}

\begin{figure*}[!tbp]
  \centering
  \begin{minipage}[b]{0.49\textwidth}
    \includegraphics[width=\textwidth]{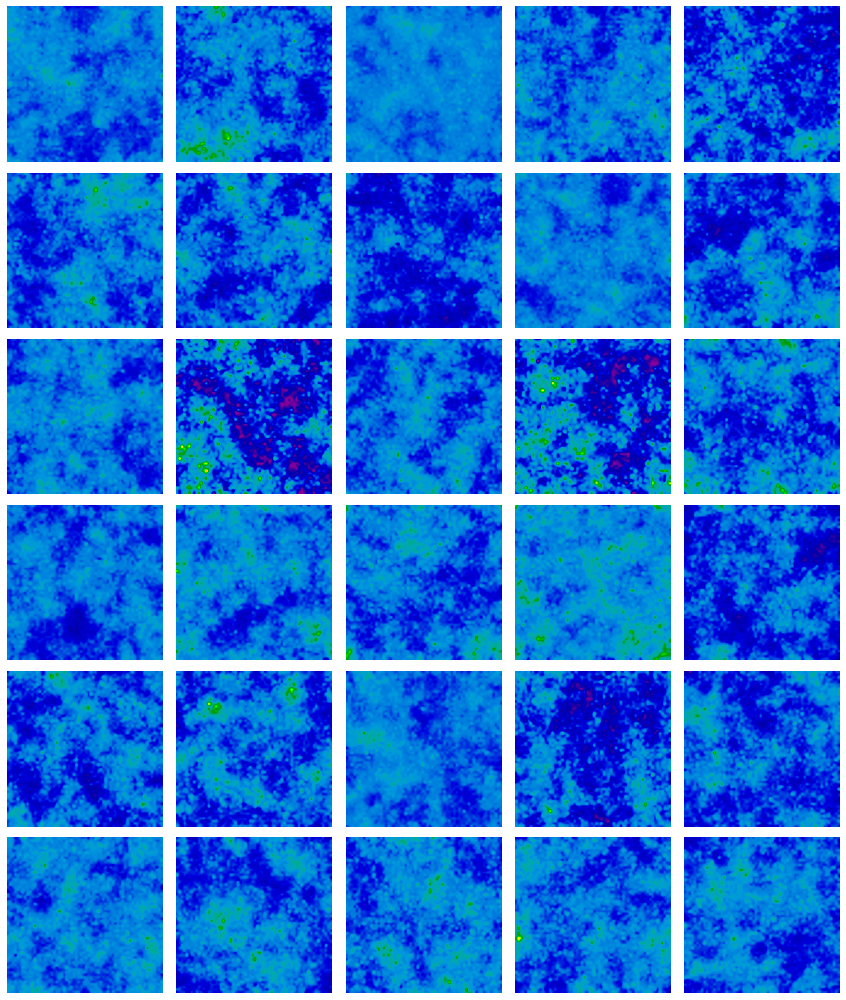}
  \end{minipage}
  \hfill
  \begin{minipage}[b]{0.49\textwidth}
    \includegraphics[width=\textwidth]{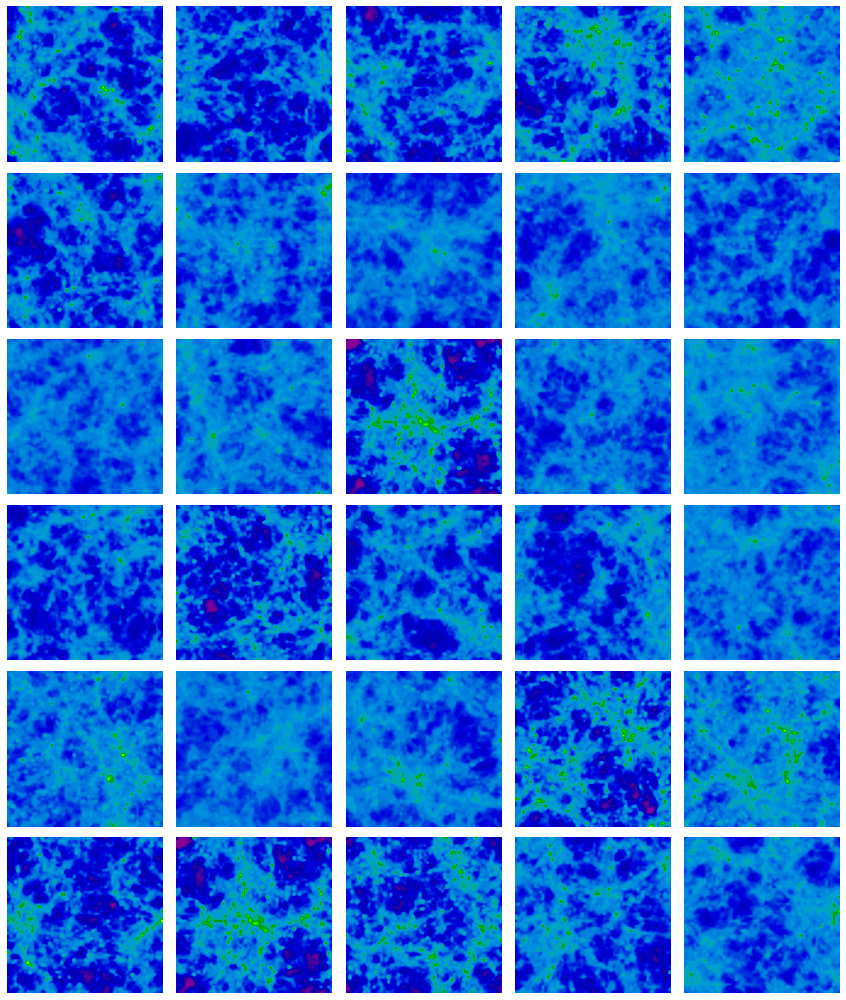}
  \end{minipage}
  \caption{Random representative examples of diverse HI maps from the CAMELS testing set (real, right) and generated using the {\it unconditional}  {\sc HIFlow} (fake, left). These maps cover an area of 25 $h^{-1}$ cMpc $\times$ 25 $h^{-1}$cMpc with 64 pixels on a side, resulting in a resolution of $\sim$ 0.4 $h^{-1}$ cMpc. The color scale in these maps show the column density range $\log_{10}$ N$_{\rm HI}$/cm$^{-2}$=14-22. \\}
   \label{fig:maps}
\end{figure*}
We choose to design {\sc HIFlow} using {\it the Masked Autoregressive Flow} (MAF), that has been shown, in~\citet{2017arXiv170507057P}, to outperform many successful density estimation methods and generative models, such as the real-valued Non Volume Preserving flow~\citep[real NVP,][]{2016arXiv160508803D}. It is worth noting that there are recently more improved models, such as Neural Spline Flows~\citep{2019arXiv190604032D} and Generative flow with Invertible 1$\times$1 Convolutions~\citep{2018arXiv180703039K}, that achieve higher performance as compared with MAF. Autoregressive models~\citep[e.g.][]{2016arXiv160502226U,2016arXiv160604934K} can be used to estimate densities and generate new examples by decomposing a joint density $p(x)$ into a product of conditionals such as $p(\textbf{x}) = \prod_{i} p(x_{i}| \textbf{x}_{1:i-1})$, which ensures that future conditional transformations are only a function of the previous values, and hence satisfies the {\it autoregressive property}. If the flow is modelled using an autoencoder (i.e. series of layers), then masking is required to remove connections between different units in different layers to preserve ordering and the autoregressive property. This approach is called Masked Autoencoder for Distribution Estimation~\citep[MADE,][]{2015arXiv150203509G}, which is the building block of the flow in MAF.  MAF increases the flexibility to learn more complex distributions by stacking several autoregressive models (MADE$_{i}, i=1...k$) into a deeper flow, where the density of the random numbers $u_{1}$ of MADE$_{1}$ is modelled with MADE$_{2}$, and those of MADE$_{2}$ with MADE$_{3}$ and so on, up to linking MADE$_{k}$ with the base (Gaussian) density. To evaluate whether MAF is able to learn the target density, the training dataset would be converted back into random numbers to test whether they represent a standard Gaussian. 

We generate HI column density (N$_{\rm HI}$) maps, along any two dimensions of size 25 $h^{-1}$ cMpc $\times$ 25 $h^{-1}$cMpc, by projecting gas particles within 5 cMpc/h columns along the third dimension from the {\sc IllustrisTNG} LH set with 64$\times$ 64 pixels, resulting in a resolution of $\sim$ $0.4$ h$^{-1}$ cMpc at z$\sim$6. These N$_{\rm HI}$ maps are basically generated by adding up the neutral hydrogen masses of gas particles within each column, and dividing by the pixel area and proton mass. This means we can generate five distinct HI maps along each of the three directions x, y and z per simulation. While these 15 maps, from each simulation, are not entirely independent as expected during {\sc HIFlow} training, we assume they are due to the small size of the training set.  In total, our dataset contains 15,000 HI maps from the 1,000 simulations of the CAMELS LH set. We use 900 simulations (or 13,500 maps) for training, and  50 simulations (or 750 maps) for validation and testing each. We convert all maps from 2-dimensional to 1-dimensional representation and transform the data to have a range from -1 to 1. Our best performing MAF, as evaluated on both validation and testing sets to show the highest averaged likelihood probability, consists of 10 autoregressive layers (10 MADE). Each MADE consists of 3 hidden layers of sizes 1024, 2048, and 4096, and each conditional is parameterized as a mixture of 10 Gaussians. Training takes approximately 30 minutes on a single Graphics processing unit (GPU). Following the terminology by~\citet{2017arXiv170507057P}, our design is called MAF MoG (10). We use the hyperbolic tangent as an activation function throughout,  Adam ~\citep{2014arXiv1412.6980K} as an optimizer to perform stochastic gradient descent via maximum likelihood, with a minibatch size of 100, a learning rate of 10$^{-4}$, a small weight decay rate of 10$^{-6}$, and early stopping is applied if no improvement is observed for the 30 consecutive epochs on the validation set. It is straightforward to extend this flow to learn the target density conditioned on a set of parameters (\textbf{y}). The conditional density would be decomposed as follows: $p(\textbf{x}|\textbf{y}) = \prod_{i} p(x_{i}| \textbf{x}_{1:i-1,\textbf{y}})$.  A visual summary of the {\it conditional} {\sc HIFlow} on cosmological parameters ($\Omega_{m}$ and $\sigma_{8}$) is shown in Figure~\ref{fig:graph}.  In this analysis, we design both unconditional and conditional {\sc HIFlow} and discuss their performance in the next section.

\begin{figure*}[!tbp]
  \centering
  \begin{minipage}[b]{0.49\textwidth}
    \includegraphics[width=\textwidth]{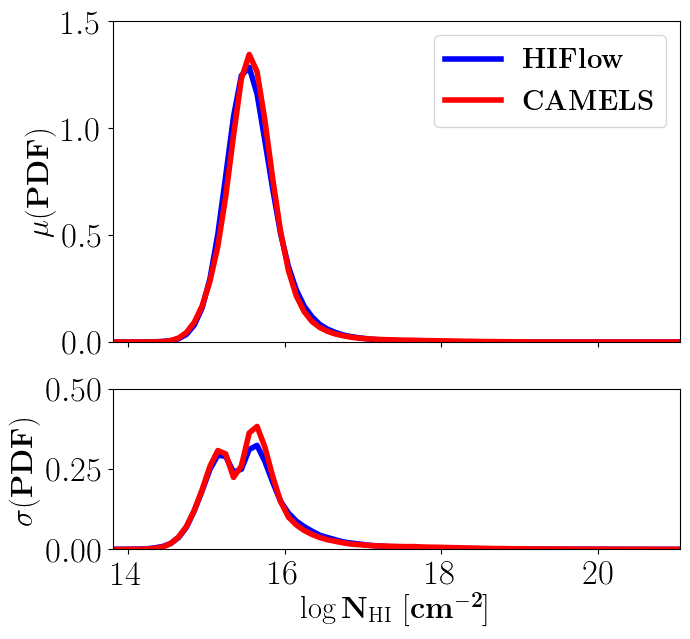}
  \end{minipage}
  \hfill
  \begin{minipage}[b]{0.49\textwidth}
    \includegraphics[width=\textwidth]{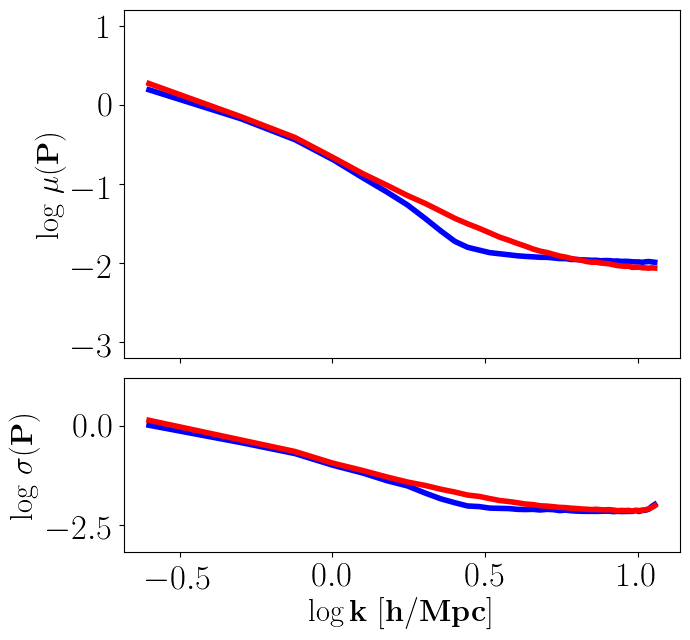}
  \end{minipage}
  \caption{Comparison between CAMELS (red) and the {\it unconditional} {\sc HIFlow} (blue), in terms of the  probability density distribution  of the HI column density pixels (PDF, left) and power spectra of the HI maps (P, right). Top and bottom panels show the average ${\mu}$(PDF) and ${\mu}$(P) and standard deviation ${\sigma}$(PDF) and ${\sigma}$(P), respectively, over the $750$ HI real maps from CAMELS testing set and the fake maps from {\sc HIFlow}. It is evident that {\sc HIFlow} recovers the expected PDF properties (average and standard deviation as a function of HI column density). While {\sc HIFlow} underpredicts the small scale power by a factor of $\sim$ 2, it predicts the expected large scale power very accurately. This effect is visible in Figure~\ref{fig:maps}.}
  \label{fig:powers}
\end{figure*}
\begin{figure*}
    \centering
    \includegraphics[scale=0.35]{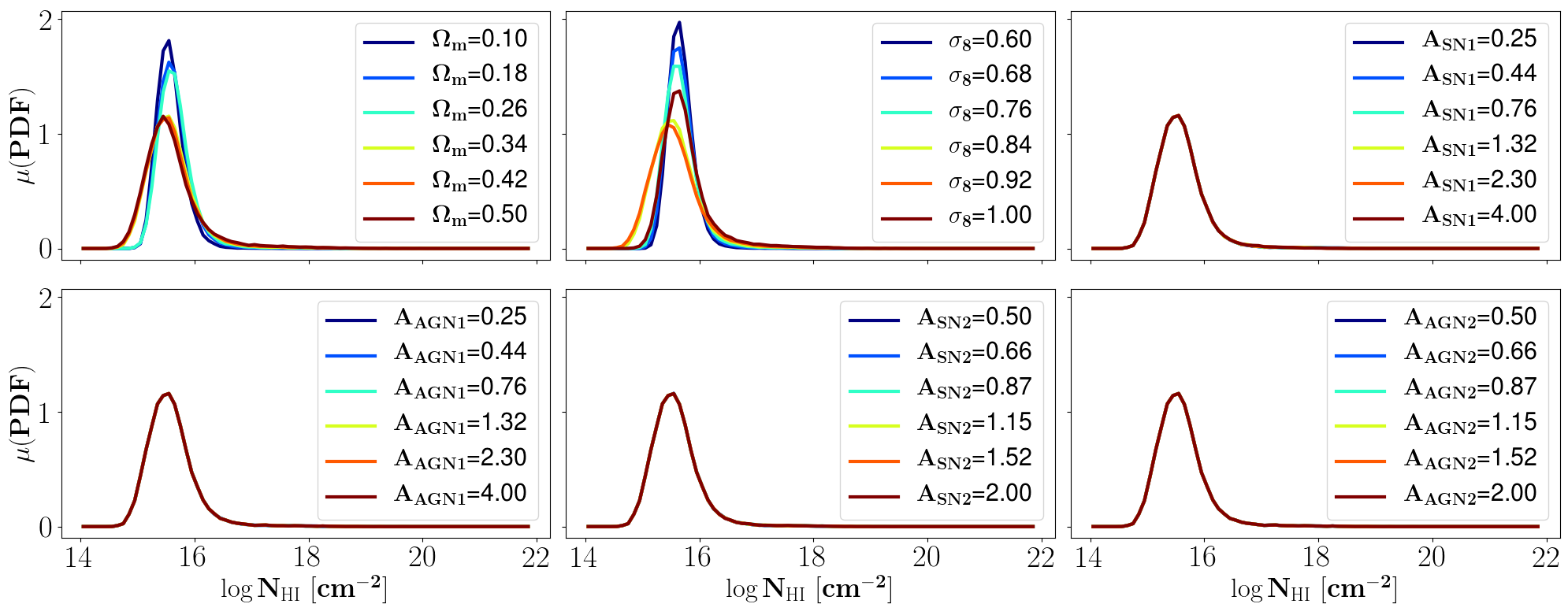}
    \caption{Impact of varying a single CAMELS parameter on the mean PDF over 15 HI maps, sharing the same parameters using the 1P set. As is evidenced by the amount of variation in the mean PDF ($\mu$(PDF)), the HI maps are mostly affected only by the cosmological parameters ($\Omega_{m}$ and $\sigma_{8}$) and not the astrophysical parameters (A$_{\rm SN1}$, A$_{\rm SN2}$, A$_{\rm AGN1}$, and A$_{\rm AGN2}$), hence we choose to condition {\sc  HIFlow} solely on cosmology. }
    \label{fig:resp}
\end{figure*}

\section{Unconditional {\sc HIFlow}} \label{sec:uhiflow}
We now test the performance of {\sc HIFlow}, without conditioning on parameters. We first show a visual comparison between the true HI maps (right) from CAMELS versus the fake HI maps generated by the unconditional {\sc HIFlow} (left) in Figure~\ref{fig:maps}. We see that the real and fake maps look very similar on large scales. On the other hand, it is clear that {\sc HIFlow} does not accurately capture the small scale features such as filaments. This is expected since the model does not explicitly incorporate the 2D structure of maps as an inductive bias. For instance, using invertible $1\times1$ convolutions would increase the model flexibility to capture the locality and reproduce the small scale features~\citep[e.g.][]{2018arXiv180703039K}. However, it is still promising to see that, without explicitly taking advantages of the 2D information, our model is able to generate diverse new examples that capture the expected large scale features reasonably well.

We now attempt to quantify the accuracy of the unconditional {\sc HIFlow} in terms of the power spectrum (P) and the 1-dimensional probability distribution functions (PDF) of the column density pixels of the generated HI maps against the CAMELS real maps. We generate 750 {\it fake} HI maps from the unconditional {\sc HIFlow} and compare them with the 750 HI maps from CAMELS testing set. We then compute the power spectra and PDFs over these 750 maps and compare the results in Figure~\ref{fig:powers}. We show the average ${\mu}$(PDF) and ${\mu}$(P) in the top panels and the corresponding standard deviation (${\sigma}$(PDF) and ${\sigma}$(P)) in the bottom panels. Comparing the average and standard deviation of the PDFs and power spectra, we see that the {\sc HIFlow} is able to reproduce CAMELS in the column density range $N_{\rm HI}\sim10^{14-21}$cm$^{-2}$, and the power spectrum within a factor of $\sim$ 2. The {\sc HIFlow} is able to recover the large scale power at $<$ 1.5 h/Mpc in wavenumber with a high accuracy.  The disagreement on small scales (around a wavenumber of 3 h/Mpc) is expected since the model only sees flattened maps during training, and hence it would be more challenging to model the highly non-linear small scales without additional information such as the 2D structure. This effect can be clearly seen in Figure~\ref{fig:maps}. Nevertheless for studies that focus on large scales, the {\sc HIflow} still is an accurate efficient tool for generating new diverse examples of the HI maps.

\begin{figure*}
    \centering
    \includegraphics[scale=0.38]{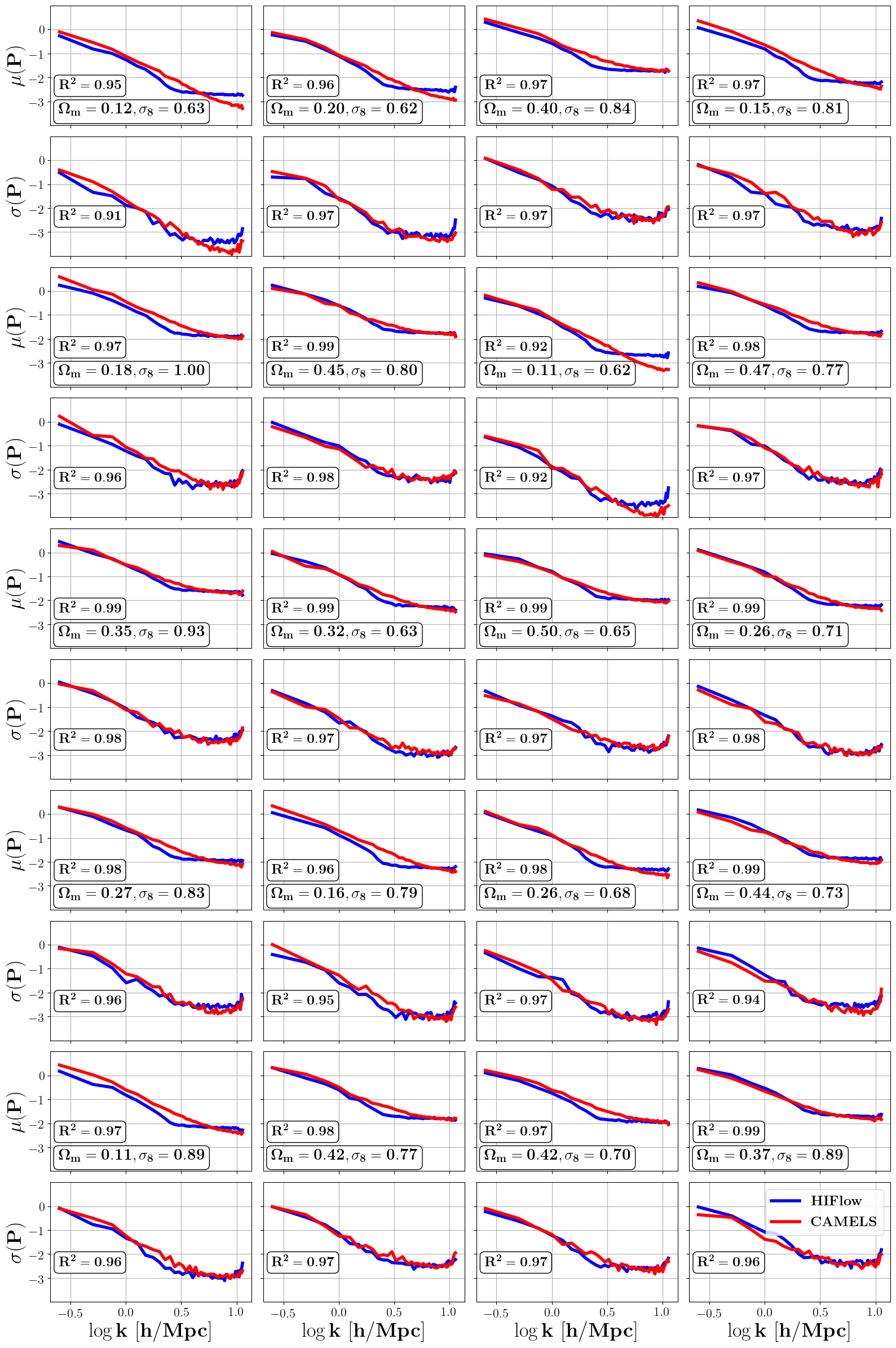}
    \caption{Comparison between the {\it conditional} {\sc HIFlow} (blue) and CAMELS (red) for randomly selected values of $\Omega_{m}$ and $\sigma_{8}$ from the testing set. The mean and standard deviation power ${\mu}$(P) and ${\sigma}$(P) are computed over the 15 maps. The $\sigma$(P) panels share the same cosmology as their immediate top $\mu$(P) panels.  In all case, the {\sc HIFlow} is able to reproduce CAMELS within a factor of $\leq$ 2, scoring a very high $R^{2} >$  90\%.}
    \label{fig:pk_all}
\end{figure*}

\begin{figure*}
    \centering
    \includegraphics[scale=0.55]{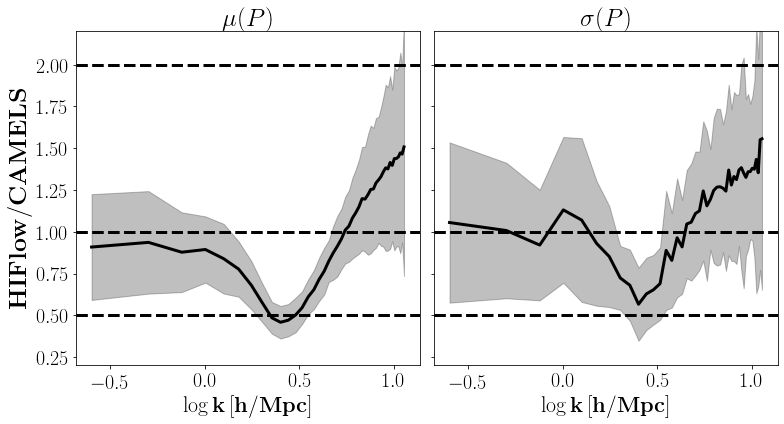}
    \caption{Ratio between the {\it conditional} {\sc HIFlow} and CAMELS of the mean  (left) and standard deviation (right) power over all the testing set as quoted in the legend. Solid lines show the average and shaded areas reflect the standard deviation over all the prior range.  {\sc HIFlow} predicts the large scale with a higher accuracy than the small scale power. In all cases, the {\sc HIFlow} is able to predict the true mean and standard deviation power within a factor of $\lesssim$ 2.  }
    \label{fig:frac}
\end{figure*}

\begin{figure*}
    \centering
    \includegraphics[scale=0.55]{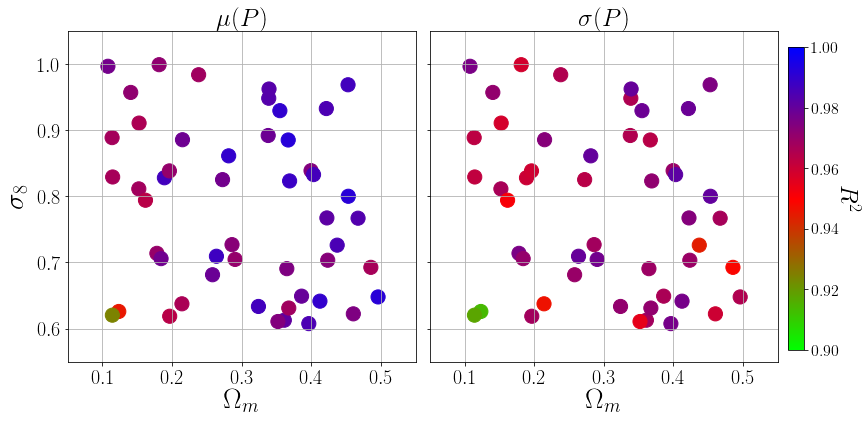}
    \caption{Visual summary of the coefficient of determination $R^{2}$ between CAMELS and {\sc HIflow} as a function of cosmology for all the testing set (50 simulations). The $R^{2}$ values for the average power ($\mu$(P)) and standard deviation power ($\sigma$(P)) are shown in the left and right panel, respectively as indicated by subtitles. The $R^{2}$ values are higher for $\mu$(P) than $\sigma$(P), but nevertheless, in all cases, there is strong correlation between the CAMELS and {\sc HIflow} with $R^{2}>$  90\%.  }
    \label{fig:r2}
\end{figure*}

\section{Conditional {\sc HIFlow}} \label{sec:chiflow}

We now focus on our other aim, which is to learn the HI maps conditioned on parameters. We first attempt to identify which parameters strongly affect the HI distribution. To do so, we make use of the CAMELS 1P set, in which a single parameter is changed at a time while keeping other parameters fixed. We generate maps from the 1P set, compute their average PDFs (${\mu}$(PDF)) and show the results in Figure~\ref{fig:resp}. We find that the stellar ($A_{\rm SN1, SN2}$) and AGN ($A_{\rm AGN1, AGN2}$) feedback parameters have no impact on the PDF of the HI maps at z$\sim$6. This might be due to the fact that at these early epochs there are fewer galaxies and AGN for their feedback to make a noticeable impact on HI maps on 25/h Mpc scales. On the other hand, the HI distribution is quite sensitive to the variation in the cosmological parameters ($\Omega_{m}, \sigma_{8}$), and hence we choose to condition our {\sc HIflow} only on the cosmological parameters, while marginlizing over the astrophysical parameters. In fact, we have initially conditioned {\sc HIFlow} on all six parameters, and found worse performance (i.e. lower averaged likelihood probability over all testing set). This is due to the degeneracies between feedback parameters as seen in Figure~\ref{fig:resp}. We will come to this point later in \S\ref{sec:inf_com}. We next re-train {\sc HIFlow}, using the same architecture, to learn the following conditional density $p( \textrm{HI}| \Omega_{m}, \sigma_{8})$ as described earlier in \S\ref{sec:maf}.

Figure~\ref{fig:pk_all} shows a comparison between the {\it conditional} {\sc HIFlow} (blue) and CAMELS (red) for randomly selected values of $\Omega_{m}$ and $\sigma_{8}$ from the testing set in terms of the power spectrum. Because we extract 15 maps from each CAMELS simulation, we generate 15 new maps from the {\it conditional} {\sc HIFlow} using the same cosmology. Matching the number of samples (15) is crucial for a consistent comparison, since smaller variance is expected for a larger number of samples (i.e. $\sigma \rightarrow \sigma / \sqrt{N}$). We then compare them using the mean and standard deviation powers ${\mu}$(P) and ${\sigma}$(P) over all 15 maps. We quote the coefficient of determination $R^{2}$ in all panels to quantify the correlation between CAMELS and {\sc HIflow} at the level of $\mu$(P) and $\sigma$(P) for the same set of the cosmological parameters. The $R^{2}$ is defined as follows:
\begin{equation}
    R^{2} = 1 - \frac{\sum_{i} (\rm X_{i} - \rm Y_{i})^{2}}{\sum_{i} (\rm X_{i} - \overline{X})^{2}}\, ,
\end{equation}

 where $X_{i}\equiv (\mu(P_{\rm CAMELS,i})$, $\sigma(P_{\rm CAMELS,i})$) and $Y_{i}\equiv (\mu(P_{\rm HIFlow,i}$), $\sigma(P_{\rm HIFlow,i}$)). The $R^{2}$ measures the fraction of the total variance within CAMELS that {\sc HIFlow} can explain. It is worth noting that the CAMELS maps are generated using six parameters, while the conditional {\sc HIFlow} is based on two parameters. Hence, these other four parameters might provide additional source of variance.

We see that in all cases, there is an impressive agreement between the {\sc HIFlow} and CAMELS in terms of the average and standard deviation powers ($\mu$(P) and $\sigma$(P)) for various set of cosmology, achieving $R^{2}> $90\%.
In Figure~\ref{fig:frac}, we show the ratio between the {\it conditional} {\sc HIFlow} and CAMELS of the mean (left) and standard deviation (right) power spectrum overall the testing set. The solid lines show the average and shaded area show the standard deviation of the ratios, whereas dashed lines show several reference lines for the perfect match and two times more/less than the true powers as produced by CAMELS. In all cases and overall the prior range, 
the {\sc HIFlow} is able to reproduce CAMELS within a factor of $\lesssim$ 2, depending on the wavenumber.  As seen before, the larger discrepancies on small scales is expected since the model does not explicitly include the expected structure of maps as an inductive bias.
\begin{figure*}
    \centering
    \includegraphics[scale=0.5]{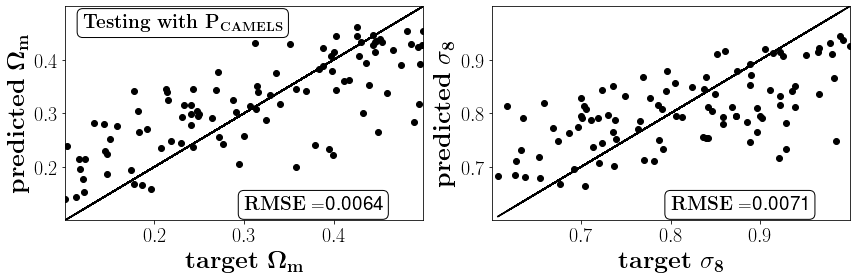}
    \includegraphics[scale=0.5]{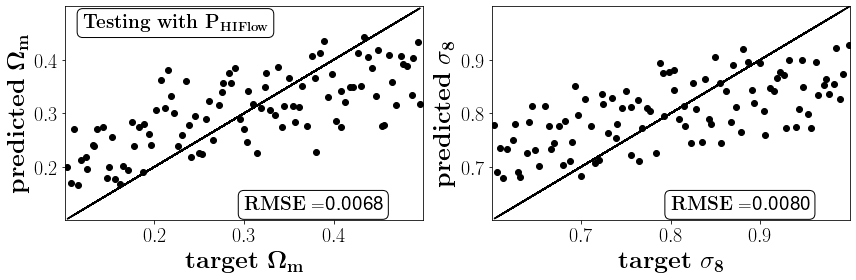}
    \caption{Correlation between the target and predicted cosmology using a simple Multi-layer perceptron trained on HI power spectra from CAMELS. Results of testing with the power spectra from CAMELS ($\rm P_{CAMELS}$) and {\sc HIFlow} ($\rm P_{HIFlow}$)  are shown in the top and bottom panels, respectively. Solid black lines represent the identity line (target versus target). While trained on CAMELS, the scatter predicted by testing with {\sc HIFlow} is similar to the target scatter produced by testing with CAMELS as quoted by the RMSE values. This shows that the differences seen in the power spectra between CAMELS and {\sc HIFlow} (see Fig~\ref{fig:pk_all},\ref{fig:frac}, \ref{fig:r2}) have a minimal impact on parameter recovery.  }
    \label{fig:ann}
\end{figure*}
We finally present a visual summary of $R^{2}$ for the testing set in Figure~\ref{fig:r2} as a function of cosmology. The $R^{2}$ values for the average power ($\mu$(P)) and standard deviation power ($\sigma$(P)) are shown in the left and right panel, respectively as quoted in subtitles. The $R^{2}$ values are higher (darker) for the $\mu$(P) than $\sigma$(P). This is expected since it is generally easier for models to reproduce the average behaviour than higher-order statistics (e.g. the variance). Nevertheless, in all cases, there is strong correlation between the CAMELS and {\sc HIFlow} with $R^{2}>$  90\%. This indicates that {\sc HIFlow} is able to explain at least 90\% of the total variance of the CAMELS' power spectra data, which is quite promising due to the small size of the training dataset, and model simplicity. It is worth nothing that we have re-trained the conditional {\sc HIFlow} several times and the results have always been similar. Hence, we do not expect an additional source of variance due to the random initialization of the network parameters.

To test whether the differences seen at the power spectrum level might impact inference and parameter recovery, we train a simple Mulit-layer perceptron on HI power spectra from CAMELS and attempt to quantify the accuracy using testing samples from CAMELS versus {\sc HIFlow} as shown in Figure~\ref{fig:ann}. The best performing Multi-layer perceptron includes 4 layers with 10 neurons each using a learning rate of  0.001. The results of testing with power spectra from CAMELS and {\sc HIFlow} are shown in the top and bottom panels in Fig~\ref{fig:ann}, respectively. The solid lines show the identify line (target versus target), and the distance to the identity lines is quantified with the Root Mean Square Error (RMSE) as quoted in the panels. We see that RMSE values and the scatter are relatively the same whether testing with data from CAMELS or {\sc HIFlow}. This indicates that the differences seen between CAMELS and {\sc HIFlow} power spectra (see Fig~\ref{fig:pk_all},\ref{fig:frac}, \ref{fig:r2}) would have a minimal impact on parameters recovery and inference.

\begin{figure*}
    \centering
    \includegraphics[scale=0.55]{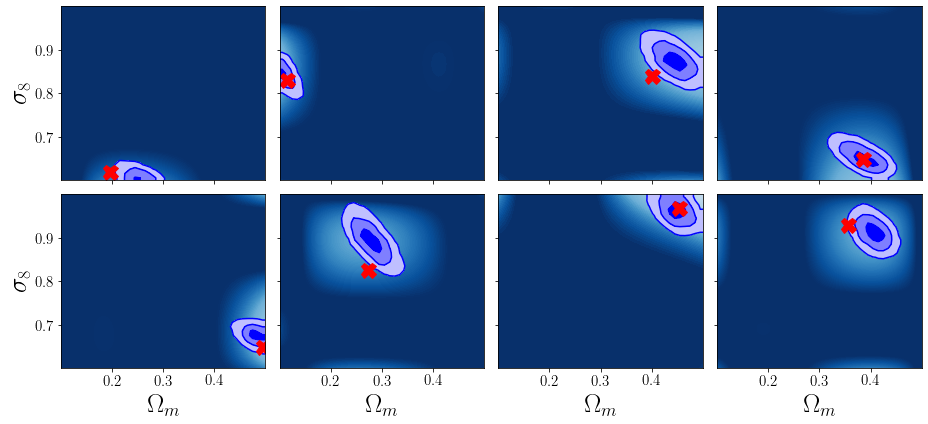}
    \caption{ Several examples of posterior distributions from {\sc HIFlow} for a random set of observed parameters (red cross). The contours represent the 1-3 sigma levels of the posterior.  The blue regions of the posterior indicate low probability values and the lighter blue show the high probability regions, where the contours reside. At different parts of the prior range, the {\sc HIFlow} is able to recover the observed HI maps within 3 sigma level. By inverting the flow, this is an illustrative example of how {\sc HIFlow} can be used to perform efficient and powerfull parameter inference.}
    \label{fig:cont_inf}
\end{figure*}

\begin{figure*}
    \centering
    \includegraphics[scale=0.55]{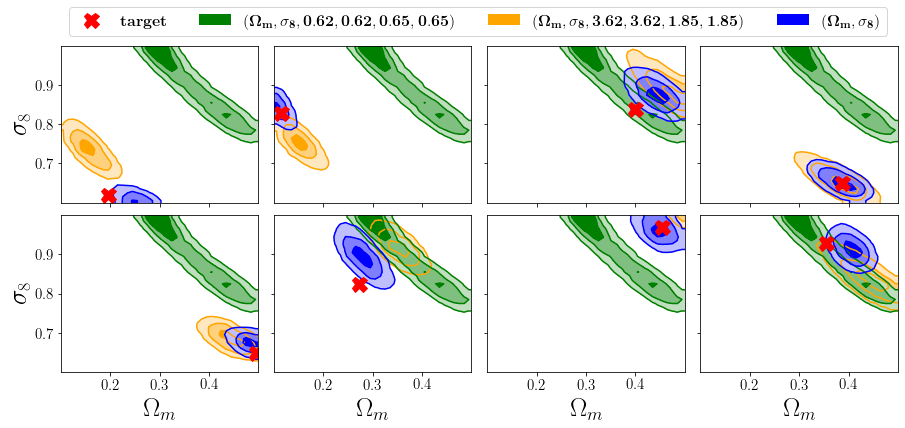}
    \caption{Similar to Figure~\ref{fig:cont_inf}, this figure shows a comparison between conditioning {\sc HIFlow} only on the cosmological parameters (blue) versus on all six parameters (green, orange) as quoted in the legend above panels ($\Omega_{m}, \sigma_{8}, A_{\rm SN1}, A_{\rm AGN1}, A_{\rm SN2}, A_{AG2}$). In all cases, we see that conditioning solely on cosmology is able to recover the observed {\sc HIFlow} maps (red cross). As seen in Figure~\ref{fig:resp}, the astrophysical parameters do not impact the HI distribution, and hence they add degeneracies which lead to posteriors far away from the target. This shows that the conditional {\sc HIFlow} on cosmology is able to successfully marginalize over the astrophysics at the field level, regardless of the feedback strength.}
    \label{fig:cont_inf_com}
\end{figure*}

\section{Inferring cosmology with {\sc HIFlow}}\label{sec:inf}
In the previous section, we have performed a comparison between maps generated from {\sc HIFlow} versus CAMELS, and found that the {\sc HIFlow}, for a given cosmology, is able to generate maps with power spectra that are in a good agreement with those expected from CAMELS in terms of the mean and standard deviation. We now focus on how {\sc HIFlow} can be used to perform parameter inference at the field level by inverting the flow. 

The advantage of normalizing flows lies in their ability to learn a tractable likelihood which can be converted to posterior distribution. In our case, since all priors are uniform, the learned likelihood is essentially equivalent to the posterior. To estimate the posterior for a given observed set of parameters as seen in Figure~\ref{fig:cont_inf}, we adopt the following approach. We first evaluate the probability for a random set of observed parameters ($\Omega_{m}, \sigma_{8}$) given their observed HI map from CAMELS (red cross symbols). We then create a uniform grid of 10,000\footnote{We have checked the results using 1,000 and 10,000 grid points and found similar posterior.} set of parameters over the whole prior range, and evaluate their probabilities given the same observed HI map. We directly use the 10,000 probabilities to estimate the 1,2, and 3 sigma levels as seen in the blue contours. To visualize the probability distribution over the whole prior range, we use the Smooth bivariate spline approximation to interpolate between probabilities to create the light/dark blue background which corresponds to the high/low probability regions given the observed cosmology (red cross) as seen in~Figure~\ref{fig:cont_inf}. In all cases, the red cross symbols from different parts in the prior range are always within the 1-3 sigma of the posterior, indicating that {\sc HIFlow} is able to recover the correct cosmologies for the selected observed HI maps. The contours are quite tight, and hence {\sc HIFlow} has the ability to exclude large part of the parameter space, and narrow down the broad range of possible scenarios. The well known negative correlation between $\Omega_{m}$ and $\sigma_{8}$ is also seen in all panels. This figure illustrates an example of how to perform inference using {\sc HIFlow} for an observed {\sc HI map}.

\section{Marginalizing over astrophysics with {\sc HIFlow}} \label{sec:inf_com}
In Figure~\ref{fig:resp}, we have shown that HI column density distribution is insensitive to the variations in the astrophysical parameters. We now turn our attention to answering the question whether performing inference with {\sc HIFlow} would be affected by the variations in these astrophysical parameters. In other words, {\it could the {\sc HIFlow} conditioned only on cosmology ($\Omega_{m}, \sigma_{8}$) marginalize over the astrophysics (the other four steller and AGN feedback parameters)?} 

To answer this question, we now condition the {\sc HIFlow} on all six parameters. We then test whether the correct cosmology can be recovered while varying the other four astrophysical parameters. We select two set the four parameters ($A_{\rm SN1}, A_{\rm AGN1}, A_{\rm SN2}, A_{AG2}$) that correspond to low (0.62,0.62,0.65,0.65) and high (3.62,3.62, 1.85, 1.85) feedback strengths, as an example. We then repeat the same inference exercise for the same observed HI maps in Figure~\ref{fig:cont_inf}, and show the results in Figure~\ref{fig:cont_inf_com}. The green and orange contours are obtained using the conditional {\sc HIFlow} on all six parameters by setting the feedback parameters to low and high strength levels as defined above, respectively. The blue contours are the same as in Figure~\ref{fig:cont_inf} that are obtained using the conditional {\sc HIFlow} only on cosmology and the red cross show the cosmological parameters for the selected observed HI maps. For all the selected observed parameters at different regions in the prior range,
the conditional {\sc HIFlow} on cosmology produces posteriors that show high accuracy in parameter recovery, since all observed parameters (red cross) are within the 3-sigma level.  On the other hand, conditioning on all six parameters clearly fails to recover the observed parameters for different feedback strengths (green, orange). This is expected since the column density distribution insensitive to the variations in the feedback parameters (see Figure~\ref{fig:resp}). The contours by conditioning only on cosmology (blue) are smaller than others (green, orange). This shows the ability of the model to exclude larger part of the parameter space, and narrow the range of possible models. For all contours, the anti-correlation between $\Omega_{m}$ and $\sigma_{8}$ exists. This figure illustrates the ability of the conditional {\sc HIFlow} on cosmology to successfully marginalize over the astrophysical parameters at the field level, regardless of the strengths of the stellar and AGN feedback.

\section{concluding remarks} \label{sec:con}
We have presented {\sc HIFlow}, an efficient and fast generative model of HI maps at $z\sim6$ from the CAMELS simulations. This new tool is designed using Masked Autoregressive Flows (MAF), which is a class of normalizing flows. The MAF used here is a stack of 10 Masked Autoencoders for Density Estimation, following closely the initial implementation by~\citet{2017arXiv170507057P}. We have trained {\sc HIFlow} on 64$\times$64 HI maps generated from the {\sc IllustrisTNG} LH set at $z\sim6$ to learn the conditional density $p({\rm HI}, \Omega_{m}, \sigma_{8})$. \\

Our key findings can be summarized as follows:
\begin{itemize}
    \item The unconditional {\sc HIFlow} is able to reproduce CAMELS HI maps in the column density range $N_{\rm HI}\sim10^{14-21}$cm$^{-2}$, and power spectrum within a factor of $\leq$ 2. While the model does not incorporate the 2D structure of maps as an inductive bias, the large scale power at $k < 1.5$ h/Mpc is recovered with a high accuracy (see Figure~\ref{fig:powers}).
    
    \item While the dependence on stellar and AGN feedback is weak, the statistical properties of the HI distributions are highly sensitive to the cosmological parameters at high redshift $z\sim6$ (see Figure~\ref{fig:resp}).
    
    \item The conditional {\sc HIFlow} on cosmological parameters (generating maps from parameters) accurately predicts the correct average and standard deviation power spectra as obtained by CAMELS  within a factor of $\leq$ 2, scoring $R^{2}>$ 90\%. (see Figure~\ref{fig:pk_all}, Figure~\ref{fig:frac}, and Figure~\ref{fig:r2}). These slight differences between CAMELS and {\sc HIFlow} do not impact cosmological inference performed using the HI power spectra (see Figure~\ref{fig:ann}).
    
    \item The {\sc HIFlow} is successfully able to perform efficient cosmological inference at the field level while marginalizing over astrophysics, regardless of the strength of stellar and AGN feedback (see Figure~\ref{fig:cont_inf} and Figure~\ref{fig:cont_inf_com}).
    
\end{itemize}

While trained on {\sc IllustrisTNG}, the same architecture as used in {\sc HIFlow} can be used to train on other state-of-the-art hydrodynamic simulations, such as  {\sc Simba}, to generate HI maps or other morphologically similar maps, and perform efficient parameter inference at the field level.  

One explanation for the good agreement between {\sc HIFlow} and CAMELS is that, at high redshift, the HI distribution is smoother and only sensitive to the cosmological parameters with no influence from the stellar and AGN feedback, as seen in Figure~\ref{fig:resp}.  For this reason, a simple normalizing flow model is able to learn the HI distribution very well by observing only the 1-dimensional representation of the maps (i.e. flattened maps). It is expected that at lower redshifts, when the stellar and AGN feedback is much stronger, a more complex architecture might be needed. It is also worthwhile noting that the HI maps are sensitive to the UV background, and hence we do not expect {\sc HIFlow} to agree with models that employ stronger/weaker UV background than what {\sc IllustrisTNG} implements at $z\sim 6$.

In addition, CAMELS employs a homogeneous ionizing background~\citep[e.g.][]{2012ApJ...746..125H}, and hence the Universe is fully ionized by z$\sim$6. If instead we used an inhomogeneous background, by modelling radiative transfer~\citep[e.g. see ][]{2021arXiv210903840H,2019MNRAS.489.5594M} either on-the-fly or in post-processing, then the HI maps would contain the non-linear morphology of ionized bubbles, which might be challenging to model with the current design of {\sc HIFlow}. In this case, advanced architectures, such as the Neural Spline Flows~\citep{2019arXiv190604032D}, Generative Flow with Invertible 1$\times$1 Convolutions~\citep[{\sc Glow},][]{2018arXiv180703039K} or the Vector Quantized Variational Autoencoder~\citep[VQ-VAE,][]{2019arXiv190600446R}, might be needed. We leave investigating more complex architectures with more complex data sets to future work. 

{\sc HIFlow} enables many applications including testing power spectral pipelines of HI surveys and assisting in computing statistical properties that require many field samples, such as the covariance matrix of HI maps or their summary statistics. {\sc HIFlow} is an initial step towards studying the non-Gaussian nature of HI maps, performing a more efficient parameter inference, powerful HI forecasting for future large scale and intensity mapping surveys, and thereby maximizing the scientific return of observations by the next generation of facilities, such as Roman and SKA. Natural next steps would include improving the model design by incorporating the data structure as an inductive bias, and  learning simultaneously different emission line intensity maps (e.g. CII, CO, Ly$\alpha$) to enable joint analysis and provide accurate predictions for future multi-wavelength and multi-messenger surveys.

\section*{Acknowledgements} 
The authors acknowledge very helpful and extensive discussions with George Papamakarios (DeepMind), which have improved the paper significantly. The authors also acknowledge comments provided by Dylan Nelson, Kaze Wong, Adrian Price-Whelan, Dan Foreman-Mackey, and Wolfgang Kerzendorf. SH, FVN, BW, DNS, DAA, SG, BLB, and RSS  acknowledge support provided by Simons Foundation. Analysis is performed at the Iron Cluster in the Flatiron Institute and the Popeye-Simons System at the San Diego Supercomputing Centre. This work also used the Extreme Science and Engineering Discovery Environment (XSEDE), which is supported by National Science Foundation grant number ACI-1548562, and computational resources (Bridges) provided through the allocation AST190009.  FVN was supported by funding from the WFIRST program through NNG26PJ30C and NNN12AA01C. DAA was supported in part by NSF grants AST-2009687 and AST-2108944.  GLB was supported in part by NSF grants OAC-1835509, AST-2108470. SA acknowledges financial support from the {\it South African Radio Astronomy Observatory} (SARAO)

\bibliography{sample631}{}
\bibliographystyle{aasjournal}



\end{document}